\documentclass[%
aip,
amsmath,amssymb,
reprint,%
author-year,%
floatfix,nofootinbib
]{revtex4-2}

\usepackage{graphicx,array,booktabs,rotating}
\usepackage{dcolumn}
\usepackage{bm}
\usepackage{mathptmx}
\usepackage[english]{babel}
\usepackage{float,color,xcolor}
\usepackage{hyperref}
\usepackage{etoolbox} 
\usepackage{tabularx,tabulary}
\usepackage[capitalize]{cleveref}
\usepackage{times}
\usepackage{multirow}
\usepackage{textcomp,fontenc}
\hypersetup{
	colorlinks,
	linkcolor={blue!100!black},
	citecolor={blue!100!black},
	urlcolor={blue!100!black}
}
\newcommand{\PRLsep}{\noindent\makebox[\linewidth]{\resizebox{0.3333\linewidth}{1pt}{$\bullet$}}\bigskip}
\makeatletter
\renewcommand*{\fnum@figure}{{\normalfont\bfseries \figurename~\thefigure}}
\renewcommand*{\@caption@fignum@sep}{\textbf{. }}
\renewcommand*{\fnum@table}{{\normalfont\bfseries \tablename~\thetable}}
\renewcommand*{\@caption@fignum@sep}{\textbf{. }}
\makeatother
\newcommand{\sk}[1] {\textcolor{black}{#1}}

\usepackage{color,float,chngcntr,xr,titlesec}
\titleformat{\subsubsection}[runin]
{\sffamily\small\bfseries}{\thesubsubsection.}{0.25em}{}[:]
\titlespacing{\subsubsection}{0pc}{1.5ex plus .1ex minus .2ex}{0.25em}
\usepackage[page]{totalcount}
\usepackage{etoolbox,fancyhdr,xcolor}%
\makeatletter
\patchcmd{\NAT@citex}
{\@citea\NAT@hyper@{%
		\NAT@nmfmt{\NAT@nm}%
		\hyper@natlinkbreak{\NAT@aysep\NAT@spacechar}{\@citeb\@extra@b@citeb}%
		\NAT@date}}
{\@citea\NAT@nmfmt{\NAT@nm}%
	\NAT@aysep\NAT@spacechar\NAT@hyper@{\NAT@date}}{}{}
\patchcmd{\NAT@citex}
{\@citea\NAT@hyper@{%
		\NAT@nmfmt{\NAT@nm}%
		\hyper@natlinkbreak{\NAT@spacechar\NAT@@open\if*#1*\else#1\NAT@spacechar\fi}%
		{\@citeb\@extra@b@citeb}%
		\NAT@date}}
{\@citea\NAT@nmfmt{\NAT@nm}%
	\NAT@spacechar\NAT@@open\if*#1*\else#1\NAT@spacechar\fi\NAT@hyper@{\NAT@date}}
{}{}

\makeatother
\usepackage[pagewise,mathlines]{lineno} 

\begin{document}
	\preprint{AIP/123-QED}	
	\title[\textbf{Exp. Fluids} (2021) $|$ Manuscript Accepted]{On the unsteady dynamics of partially shrouded compressible jets}

	\author{Soumya R. Nanda}
	\affiliation{Department of Aerospace Engineering, Indian Institute of Technology, Kanpur 208016, India}%
	
	\author{S. K. Karthick}
	\affiliation{Faculty of Aerospace Engineering, Technion-Israel Institute of Technology, Haifa-3200003, Israel}%
	
	\author{T. V. Krishna}
	\affiliation{Department of Aerospace Engineering, Indian Institute of Technology, Kanpur 208016, India}%
	
	\author{A. De}
	\affiliation{Department of Aerospace Engineering, Indian Institute of Technology, Kanpur 208016, India}%
	
	\author{Mohammed S. Ibrahim}
	\email{ibrahim@iitk.ac.in (Corresponding Author)}
	\affiliation{Department of Aerospace Engineering, Indian Institute of Technology, Kanpur 208016, India}%
	
	\date{Accepted on September 20, 2021}
\begin{abstract}
We experimentally investigate a partially shrouded sonic jet (a sonic free-jet shielded by a solid wall-extension on one side) exiting from a planar nozzle at two different nozzle pressure ratio ($\zeta=4$ and $5$). We experimentally show that the inherent jet unsteadiness from the shock-induced flow separation on the wall and the emitted noise in the far-field is strongly coupled through a series of experiments like high-speed schlieren, wall-static pressure, unsteady pressure spectra, and microphone measurements. The partially shrouded jet's lateral free expansion is also identified to be complicated, three-dimensional, and the produced noise is directional. The emitted acoustic pulses from the flapping-jet, the radiated noise from the shock-induced separation on the wall, and the shock-shear layer interaction on the other side of the wall are responsible for the generated acoustic disturbances. The non-uniform aeroacoustic forcing on the top and bottom portion of the partially wall-bounded jet shear layer leads to a self-sustained jet oscillation and a discrete sound emission. The vital features are identified through the proper orthogonal decomposition of high-speed schlieren images and supplemented by other measurements.
\end{abstract}


\maketitle

\section{Introduction} \label{sec:intro}
Partially shrouded compressible jet is a type of asymmetric jet \citep{Wlezien1988} that is encountered in a variety of aerospace applications involving thrust-vectoring \citep{Sutton2006}, thrust-augmentation \citep{Veen1974}, fluid-mixing \citep{Zeoli2008}, and noise suppression \citep{Viswanathan2011}. Nozzles exhausting such kind of jets are being researched under various names based on the applications like non-conventional nozzles \citep{Hiley1976}, stepped-nozzles \citep{Wei2019}, plug-nozzles \citep{Romine1998}, ramp-nozzles \citep{Berry2017b}, nozzle with aft deck \citep{Behrouzi2018} and multi-stream nozzles \citep{Stack2018}. In these kinds of nozzle flows, the exhausting jet is partially covered by similar/different fluid streams and partially bounded by a solid-wall. The nozzle's structural integrity primarily depends on the condition of whether the jet is attached or detached from the solid-wall. 

Variable nozzle operating conditions lead the jet to expand imperfectly and influence the jet separation characteristics. Flow separation from the wall is inherently an unsteady phenomenon and exerts unusual load distribution on the partial shrouds, in-particular during nozzle over-expansion. Besides, the shock-laden compressible jet emits periodic and broadband pressure pulses from screeching and turbulent flow-mixing. These phenomena are particularly prominent in pulse detonation engines (PDE), which also suffer from combustion chamber flow unsteadiness \citep{Zhu2020}. \sk{In summary, the burnt exhaust gas from the PDE's asymmetric nozzle experiences operational difficulties from both the driving flow unsteadiness and the resulting acoustic fluctuations.} All these unsteady loads contribute to the air-frame fatigue and pose a threat to the vehicle's stability.

\cite{Dosanjh1988, Das1991, Das1997} have performed a dedicated experimental and computational campaign on the jet aeroacoustics from the plug-nozzle. The perforation percentage and length of the partially bounded walls are varied, and the resulting far-field noise emission is calculated in different directions. A shorter and porous confinement passage has exhibited noise suppression of up to 3 dB during off-design operations. A computational noise prediction scheme has been proposed, which predicts noise levels for under-expanding jets within 5 dB variations except at high-frequencies. Experimental studies have further revealed that sonic jets produce higher sound intensity than the supersonic jets exhausting through contoured convergent-divergent (CD) plug-nozzle.

In the experiments of \cite{Verma_2011, Chutkey_2012, Chutkey_2017, Chutkey_2018}, planar and cluster jets from plug-nozzles are investigated by monitoring the unsteady pressure spectra on the wall for different conditions. The varying unsteady jet dynamics in the presence and absence of a base flow, side plates, and wall truncation are studied. The low-frequency high-amplitude contents in the over-expanded jet are attributed to the inherent flow unsteadiness from the shock-wave turbulent boundary layer interactions.  Oil-flow imaging is also done to visualize the three-dimensional nature of the expanding jet at different operating conditions. However, information on other discrete tones at high-frequency in the unsteady spectra is scarcely discussed.

\sk{Screeching jet produces high-frequency discrete noise. Such jets are notable to contain shock motions or shock-cell oscillations too. In general, shock or shock-cell oscillation occurs at different spectra based on the flow problem in consideration and its operating conditions. For example, the pseudo-shock oscillation in duct flow or shock-cell oscillation in a supersonic confined jet is observed between 10-100 Hz \citep{Rao2014,PethaSethuraman2021}. Shock oscillation due to flow separation ahead of protrusion or deflection is in the order of 100 Hz \citep{Clemens2014,EstruchSamper2018}. Shock related unsteadiness from an imperfectly expanded nozzle flow separation is between 500-1000 Hz \citep{Chutkey_2018,Li2019}. The unsteadiness in the last case like the plug-nozzle flow resembles generic flow physics related to the partially shrouded compressible jets.}

\cite{Wei2019} report the mitigation of jet noise arising from the under-expanded stepped nozzle. They observe an inconsistent trend in jet noise radiation and broadband shock associated noise (BSAN) about the nozzle pressure ratio variations. The generic correlations proposed for free jets in shock-cell lengths and BSAN-Strouhal numbers are reportedly not matching the semi-shrouded jets from the stepped nozzles. However, a global noise reduction and wider jet-spread are observed from the microphone measurements and high-speed schlieren imaging.

\begin{figure*}
  \centerline{\includegraphics[width=0.8\textwidth]{ 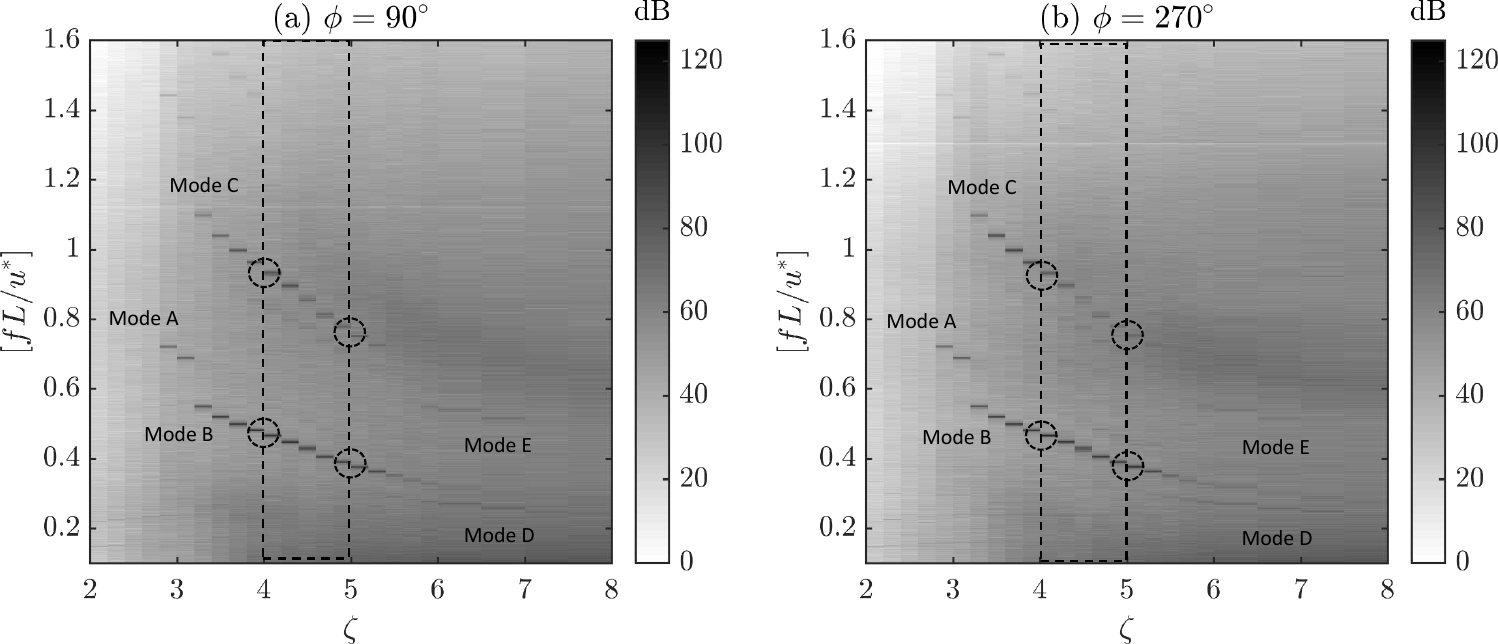}}
  \caption{\sk{Typical spectral variations obtained during the experimental campaign of \cite{Chaudhary2020} for a wide range of nozzle pressure ratio ($2\leq \zeta \leq 8$, see Table \ref{tab:flow_cond} for nomenclature clarification) at two different directions: (a) $\phi=90^\circ$, and (b) $\phi=270^\circ$ (see Figure \ref{fig:schematic} for microphone placement) for the laterally unbounded partial shrouded jet as described in Figure \ref{fig:nozzle_dimension}. The contour plots show the presence of intermediate instability mode called \textit{Mode-B} for $3.2\leq \zeta \leq 6.2$. The vertical dotted-lines represent the targeted $\zeta$ (4 and 5) within the region of dominant noise production. The dotted-circles represent the dominant screech tone observed at the respective conditions.}}
\label{fig:npr_mic}
\end{figure*}

In the series of research work by \cite{Berry2017c,Berry2017b,Berry2017a} on the jet from a nozzle with an aft deck, the distinct spatial modes associated with the oscillating jet are identified from the modal decomposition of the high-speed schlieren images (both Proper Orthogonal and Dynamic Mode Decomposition-POD/DMD). The screeching jet characteristics at a particular frequency and the corresponding spatial mode are highlighted for variety of geometrical and flow conditions.

Some of the present authors' previous work on jets from plug-nozzle \citep{Khan_2019, Chaudhary2020} have elaborately considered the influence of flow separation-related unsteadiness arising from lateral confinement. The vanishing of flow three-dimensionality and the shift of dominant spectral components near the separation point upon confinement are argued using the wall-static unsteady pressure measurements and the complementary three-dimensional RANS (Reynold Averaged Navier-Stokes) simulations. One of the major conclusions is the power reduction in the unsteady spectra, attributed to lateral confinement's flow shielding from the jet screech's noises.

\sk{Cursory microphone measurements taken during the experimental campaign of \cite{Chaudhary2020} for the unbounded partially shrouded jet show the presence of different instability modes (Figure \ref{fig:npr_mic}). The instability modes: $Mode-B$ present between $3.2\leq \zeta \leq 6.2$ ($\zeta=p_0/p_a$, ratio of stagnation and atmospheric pressure) and $Mode-C$ ($3.2\leq \zeta \leq 5.2$) in general contribute to the dominant noise production. In axisymmetric \citep{Panda1997} or elliptic \citep{EdgingtonMitchell2019} supersonic free jet, $Mode-B$ and $Mode-C$ is related to sinuous and helical jet instability. The modes themselves have a strong relation such that the frequency of the first harmonic is twice the fundamental ($f_1=2f_0$). In a rectangular jet \citep{Tam1988}, only $Mode-B$ is present and it is surprising to see a contrasting spectra in Figure \ref{fig:npr_mic} for the partially shrouded rectangular jet. The unsteady jet separation and the coupling between jet aeroacoustics are suspected to drive the jet instability. Such an investigation warranted a separate study and was not in the scope of \cite{Chaudhary2020}. Moreover, $\zeta$ at the extrema of $Mode-B$ were only considered ($\zeta=3$ and, $6$) and the studies were only from the perspective of jet dynamics and confinement influence but not in tandem with the jet aeroacoustics.}

Recently, in the works of \cite{Rao2019,Rao2020}, the extraction of flow-dominant mode and the associated spectral content from the under-sampled data are successfully demonstrated for the schlieren images of moderate exposure time and frame rate. The effect of aliased temporal spectra are also validated using the microphone measurements. The dominant energetic and dynamic modes are shown to be the same between the cases of high-exposure and short-exposure imaging at low frames-per-second (fps). Later, the coupling of acoustic modes with the flow modes are shown from the final analysis of the schlieren images from a screeching conical and elliptical jet. Similarly, flow separation in spiked-bodies and the resulted shock-related unsteadiness are identified through the dominant spatio-temporal modes from the modal analysis of high-speed shadowgraph images in \cite{Sahoo2020,Sahoo2021}.

Asides the important dynamic studies on compressible jets in the recent days, there are many other fundamental studies on partially shrouded jets. Those studies are mostly dominated by steady-state or time-averaged analysis in terms of capturing the jet morphology, identifying the separation point, and measuring the wall-static pressure. Many parametric studies are also found in the open literature involving variations in the confinement extent, base flow conditions, exhaust-nozzle contour, and the wall profile. For the brevity of literature review, those studies are not discussed in detail.


\sk{From the brief research summary, in particular from the previous work of \cite{Chaudhary2020}, following motivation is drawn. The partially bounded compressible jet's separation characteristics and the associated noise emission have not been studied in an ad-joint manner towards understanding the underlying flow physics. Especially in the dominant noise producing operating conditions where the influence of $Mode-B$ and $Mode-C$ jet instabilities are severe.} Hence, an appropriately planned experimental campaign is of utmost importance as the ensuing knowledge helps develop novel strategies to control and mitigate the adverse effects from off-design operations. Qualitative tools like the oil-flow visualization and schlieren imaging at moderate speed and quantitative tools like the modal analysis of high-speed schlieren images along with the unsteady pressure and microphone measurements are likely to help map the general jet dynamics.

\sk{In the present paper, we consider a partially shrouded sonic jet emanating from a planar-plug type nozzle at $\zeta=4$ and $5$ where the most unstable jet operating conditions are encountered.} Following are the distinct objectives of the present paper:
\begin{enumerate}
    \item {To identify the dominant unsteady frequency of the expanding jet along the wall using steady and unsteady pressure sensors.}
    \item {To capture the corresponding dominant noise emission at different directions in the far-field by studying the shrouded jet inside a semi anechoic chamber using a microphone.}
    \item {To map the resulting dominant spatiotemporal features of the shrouded jet from the modal analysis of high-speed schlieren images.}
    \item {To verify the existence of coupling between the unsteady flow separation and the far-field noise.}
\end{enumerate}

The rest of the manuscript is organized as follows. In terms of qualitative and quantitative measurements with associated uncertainties, relevant details about the experiments are elaborated in $\S$\ref{sec:expt_method}. Key results and discussions from the high-speed schlieren images including $x-t$ plots, and modal analysis, followed by the unsteady pressure, and microphone measurements are given in $\S$\ref{sec:res_disc}. Major conclusions from the experimental campaign are enlisted in $\S$\ref{sec:conclusions}.

\section{Experimental Methodology} \label{sec:expt_method}
\subsection{\sk{Blow-down facility with a semi anechoic chamber}} \label{sec:facility}

\begin{figure*}
  \centerline{\includegraphics[width=1\textwidth]{ 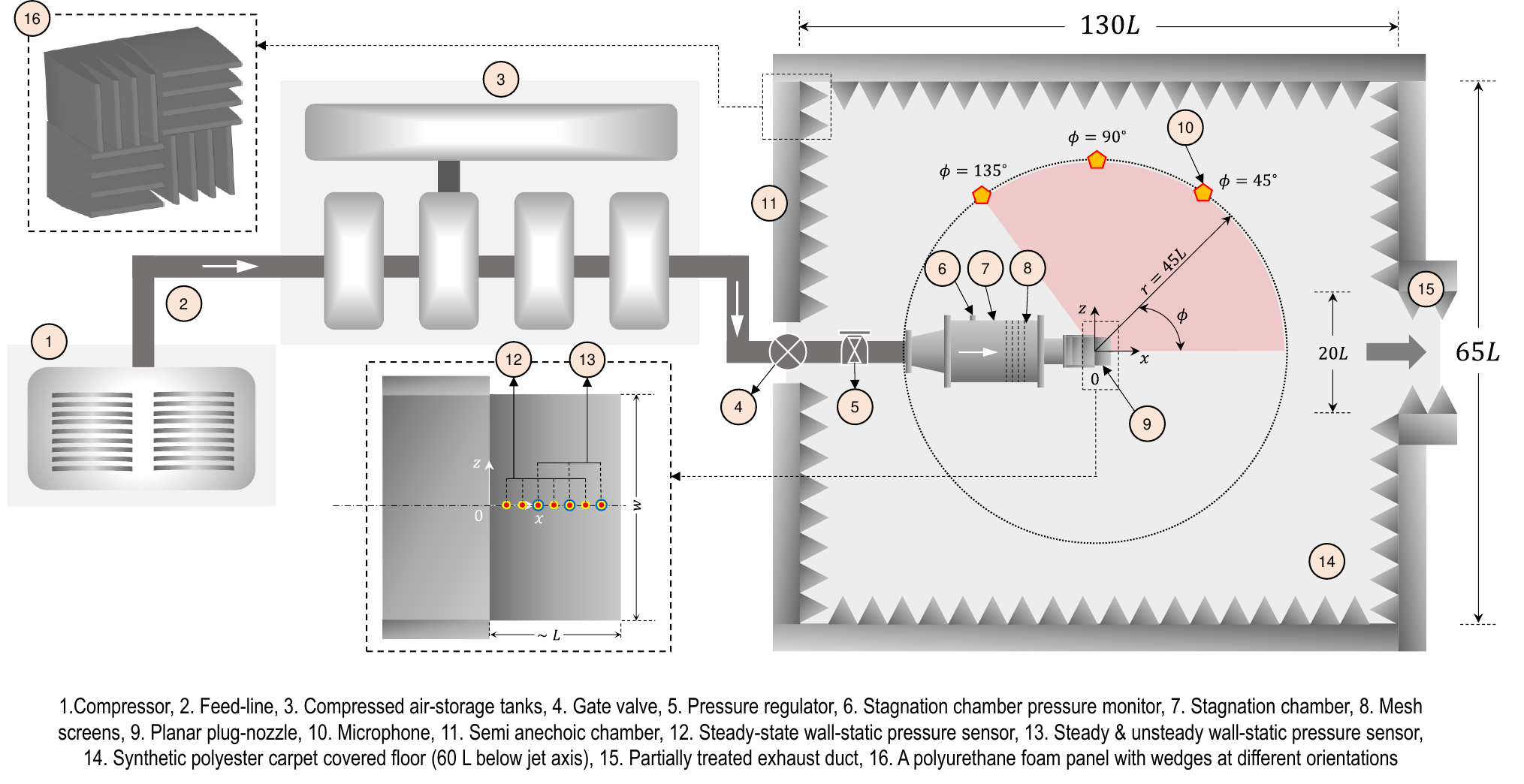}}
  \caption{\sk{A detailed schematic of the blow-down type jet-flow facility enclosed in a semi anechoic chamber shows the far-field microphone placements about the $x-z$ plane passing through the jet axis. In the snippet view, the steady and unsteady pressure sensor mounting locations on the one-sided solid wall-extension about the nozzle's exit plane that partially shrouds the sonic free-jet are also shown (Note: The schematic is not drawn to the scale).}}
\label{fig:schematic}
\end{figure*}

All the experiments are carried out at the High-speed Aerodynamics Laboratory (Aerospace Engineering, Indian Institute of Technology-Kanpur). A detailed layout of the utilized facility is given in Figure \ref{fig:schematic}. The facility is of `blow-down' type, and the jet opens into the atmosphere. A multi-stage piston-type air compressor (discharging 0.17 m$^3$/s of air at 3.5 MPa) with a dehumidifier and associated air-oil filters are used to store the clean and dry compressed air in the multiple-storage tank (maximum pressure of 2 MPa and a total volume of 85 m$^3$). A gate-valve and a pressure regulator are used in combination to operate the open-jet facility at a desired total pressure ($p_0$) measured at the stagnation chamber. The jet-facility's total temperature remains constant at an ambient temperature of $T_0=T_a=300$ K, and a maximum steady test-time of 5-10 s is achieved. A series of mesh-screens of varying size are arranged along the streamwise direction inside the stagnation chamber to minimize the flow non-uniformity with lower turbulence intensity. 

\sk{The jet facility is centrally confined by a semi anechoic chamber whose dimensions are $130L \times 65L \times 130L$ (length, width, and height, $L$=reference length) as marked in the snippet of Figure \ref{fig:schematic}. The jet axis is at an elevation of $60L$ from the ground. The selected chamber dimensions are proven to be sufficient enough to investigate jet aeroacoustics atleast in the near-field \citep{Dhamanekar2013,Baskaran2019}. Polyurethane foam panels (each of size 0.3 m $\times$ 0.3 m) made of a thick supporting base (0.15 m) with protruding wedges (base $\times$ height, 0.075m $\times$ 0.21 m) at two different panel orientations ($0^\circ$ and $90^\circ$) are used as the sound absorbing material everywhere except the floor. The chamber's basement is covered with a synthetic polyester carpet to absorb sound to certain extent and to help mounting the other devices. The jet exhaust catchment duct is only partially treated with the foam materials and the rest of the ducting is kept simple. The semi anechoic chamber's lower cut-off frequency is estimated to be around 400 Hz \citep{Sushil2020}.}

\subsection{Jet operating conditions} \label{sec:jet_op}

\begin{figure*}
  \centerline{\includegraphics[width=0.8\textwidth]{ 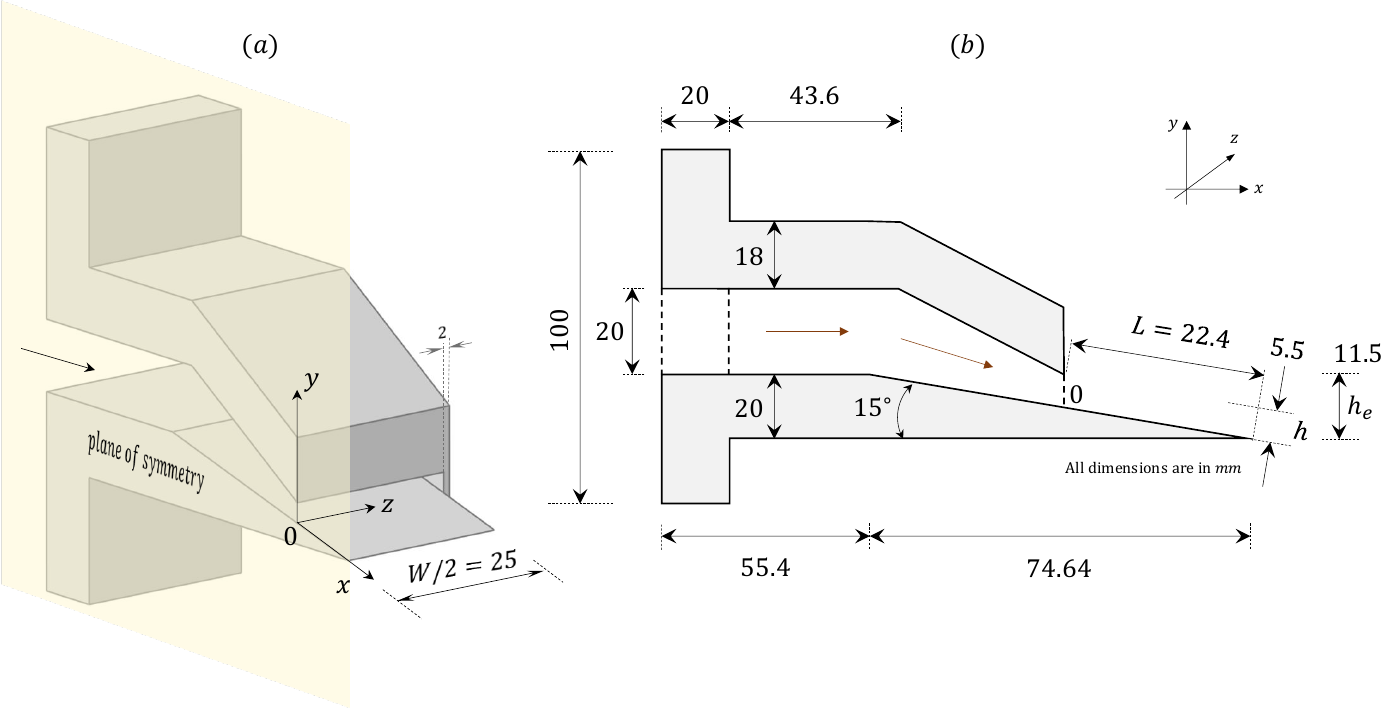}}
  \caption{(a) Three-dimensional (one-half about the plane of symmetry) and (b) two-dimensional sketch of the planar nozzle and the ramp (solid wall-extension) along with the key geometrical details and coordinate system.}
\label{fig:nozzle_dimension}
\end{figure*}

\begin{figure*}
  \centerline{\includegraphics[width=0.8\textwidth]{ 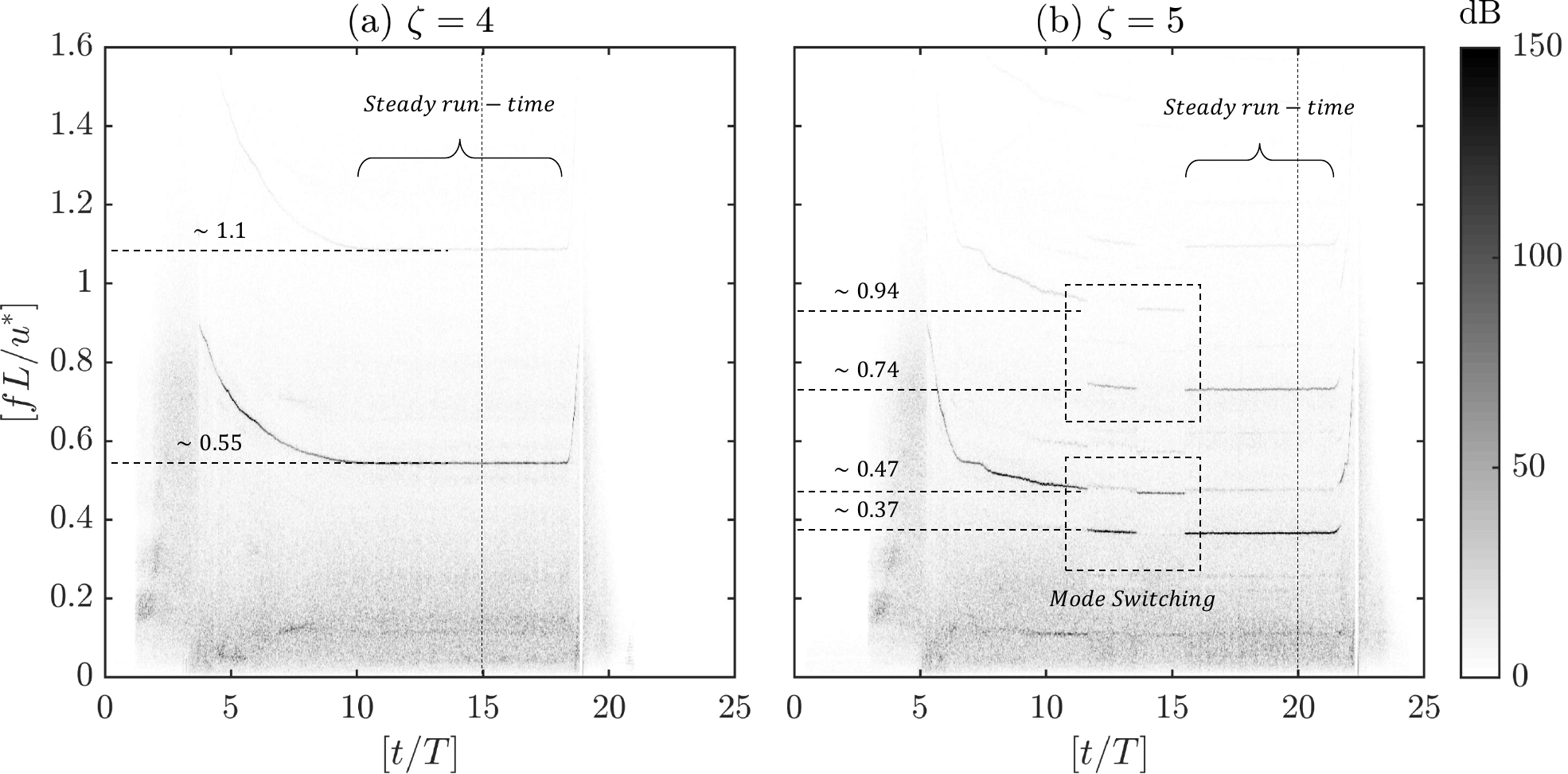}}
  \caption{A typical spectra obtained from the Shenggu\textsuperscript{\textregistered} SG-108 unidirectional microphone mounted on the oil-flow imaging camera showing the dominant frequencies during transient and steady-state operation, which match fairly well with the array-microphone measurements shown in Figure \ref{fig:spectra}-I. The vertical line marks the start-time of data acquisition for other measurements, in general. A reference time of $T=1$ s is used for non-dimensionalization. Corresponding schlieren video file is available at \href{https://youtu.be/ZN-0Ah19Pwc}{`video1.mp4'} and the animation of the microphone spectrum is available at \href{https://youtu.be/1t0hw1pHnLw}{`video2.mp4'} in the supplementary.}
\label{fig:run_time}
\end{figure*}

A rectangular planar nozzle of aspect ratio $[W/h]\approx 9$ is considered for the experiments (where $h$-nozzle height, and $W$-nozzle width in the exit plane). \sk{Two reasons are behind the selection of respective nozzle aspect ratio: 1. Rectangular plug/ramp nozzles used particularly in nozzles for improved aerodynamic efficiency are generally between 4-12 (moderate to high, \cite{Chutkey_2012,Malla2021}); 2. Limitation of the present blow-down facility in achieving a meaningful run-time to collect all the data as higher aspect ratio ($W/h \gtrsim 9$) will render short run-time and lower aspect ratio ($W/h \lesssim 5$) will not exhibit the coupled presence of $Mode-B$ and $Mode-C$ jet instabilities \citep{Zaman1996}.} The jet expands freely to the atmosphere ($p_a$) in all directions except in the bottom-side where the solid wall-extension exists whose lateral sides are unbounded. The wall-extension is like a ramp in a typical planar-plug nozzle \citep{Chaudhary2020} with a ramp-angle of $15^\circ$ as shown in Figure \ref{fig:nozzle_dimension}-a. The slant-length of the ramp is considered as the reference length ($L=22.4$ mm). The vital dimensions of the nozzle-ramp assembly are given in Figure \ref{fig:nozzle_dimension}-b along with the marked origin, flow direction, and co-ordinate system. The jet operating conditions are described in terms of the fully-expanded jet parameter \citep{Chaudhary2020,Karthick2016} in Table \ref{tab:flow_cond}. The nozzle operating conditions are changed by varying the nozzle pressure ratio ($\zeta=p_0/p_a$). Spectrograms in Figure \ref{fig:run_time} show the transient and steady-state conditions achieved during a typical run for $\zeta=4$ and $5$. More details about Figure \ref{fig:run_time} are discussed in $\S$\ref{sec:shock_acous}.

\begin{table*}
\begin{ruledtabular}
  \begin{tabular}{ccccccc}
      $\zeta=p_0/p_a$  & $u_j$ (m/s)   &   $\nu_j \times 10^{-5}$ (m$^2$/s) & $h_j$ (mm) & $M_c=0.55u_j/a_a$ & $M_j$ & $Re_j \times 10^6$\\[3pt]
      \midrule
       4 & 443.97 & 0.771 & 6.7 & 0.72 & 1.56 & 3.862\\
       5 & 471.34 & 0.684 & 7.4 & 0.76 & 1.71 & 5.104\\
  \end{tabular}
  \caption{Tabulation of jet flow quantities in terms of the fully-expanded jet Mach number ($M_j$), velocity ($u_j$), and height ($h_j$) at a constant total temperature of $T_0=T_a=300$ K. Reference values for non-dimensionalization: $L=22.4$ mm (reference length: planar-plug nozzle's ramp-wall length), $u^*=316.94$ m/s (throat velocity), $a_a=340.6$ m/s (sound propagation velocity in the ambient), and $p_r=20 \times 10^{-6}$ Pa (reference pressure). $Re_j=[u_jh_j/\nu_j]$, $\nu_j$-kinematic viscosity, $M_c$-convective Mach number.}
  \label{tab:flow_cond}
  \end{ruledtabular}
\end{table*}

\subsection{Flow visualization} \label{sec:flow_vis}

Oil flow and schlieren visualization are performed to access flow features on the ramp-wall and freestream qualitatively. Oil flow imaging \citep{Terzis2011} is done by first pasting a thin black-acrylic sheet on the ramp surface for getting a good contrast. Later, a suitable mixture of oleic acid, vacuum oil, kerosene, and titanium dioxide is prepared until a tooth-paste consistency is achieved. The model surface is primed with kerosene, and the prepared paste is sprayed almost uniformly over the surface using a brush and comb just before the run. A digital camera (Nikon\textsuperscript{\textregistered} D7000, $1280 \times 720$ pixels at 0.0744 mm/pixel with a sampling rate of $f_s$=60 Hz and an exposure time of about 17 ms) \sk{with is used to capture} the flow patterns during the run by placing it perpendicular to the ramp-wall surface. The camera sensor plane is mounted at a vertical distance of $[z/L]\sim 55$ and the imaging is done using a Nikkor\textsuperscript{\textregistered} 70-300 mm $f/4.5-6.3$ lens. The camera is also mounted with an external Shenggu\textsuperscript{\textregistered} SG-108 unidirectional microphone to measure relative sound pressure variation particularly during oil-flow imaging at $\psi=90^\circ$ (radial angle along the $xz$-plane). The microphone samples audio at $f_s$=48 kHz and has a flat-response between 30-18000 Hz with a sensitivity of (1 V/kPa at 1 kHz). 

Schlieren imaging \citep{Settles2001} is done using the `z-type' optical arrangement consisting of a pair of parabolic and planar-reflecting mirrors (200 mm diameter and 1.5 m focal length), a vertical knife-edge, a white-light LED (3 W, Holmarc\textsuperscript{\textregistered}), and a high-speed camera. A Cronos\textsuperscript{\textregistered} high-speed camera along with a Computar\textsuperscript{\textregistered} 12.5-75 mm $f/1.2$ lens is used to capture the images at two different resolution to resolve features both spatially (imaging type-I: $600 \times 300$ pixels at 0.1613 mm/pixel with $f_s$=1 kHz and 70 $\mu$s light exposure time) and temporally (imaging type-II: $336 \times 180$ pixels at 0.2083 mm/pixel with $f_s$=25.126 kHz and 50 $\mu$s light exposure time). Nearly 5000 images are recorded and 1000 images are used in batch to resolve the temporal details. Thus, the spectral resolution (and spectral range) for both types of imaging is calculated to be 1 Hz (0-500 Hz) and 25.13 Hz (0-12.563 kHz), respectively.

\subsection{Dynamic measurements} \label{sec:dynamic_meas}

\begin{figure*}
  \centerline{\includegraphics[width=0.85\textwidth]{ 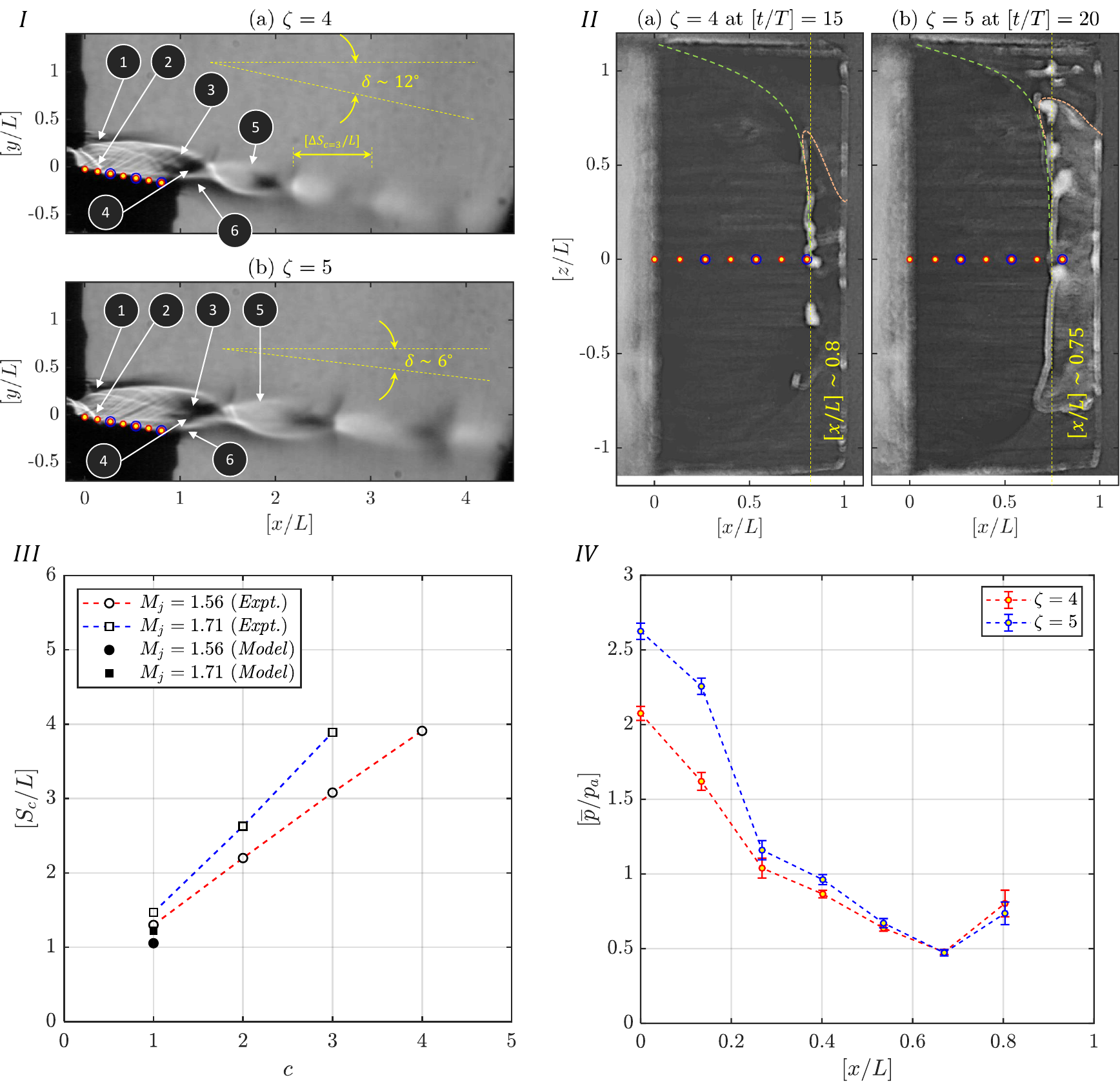}}
  \caption{I. Time-averaged schlieren images of type-I imaging showing the streamwise normalized density gradients ($\parallel{\partial{\rho}/ \partial{x}}\parallel$) highlighting the jet deflection angle ($\delta$) at $\zeta = 4,5$ (a, b). Prominent flow features: 1. free-shear layer, 2. Expansion wave, 3. Reflected compression waves, 4. Over-expansion shock, 5. Shock-cell, and 6. Separated boundary layer.  Corresponding video file is available at \href{https://youtu.be/1t0hw1pHnLw}{`video2.mp4'} in the supplementary; II. Instantaneous oil-flow imaging at $\zeta = 4,5$ (a, b) with reference to Figure \ref{fig:run_time}; III. Non-dimensional Shock-cell distance ($S_c/L$) plotted as the function of shock-cell number ($c$) for different fully expanded jet Mach number ($M_j$) corresponding to $\zeta = 4,5$. Model proposed by \citep{Norum1982} for planar jets is used to compare the first shock-cell spacing ($c=1$) as given in Eq. \ref{eq:cell_space}; IV. Non-dimensional time-averaged wall-static pressure distribution ($\overline{p}/p_a$) along the ramp (location of sensors are marked in I and II - red-outlined yellow markers are steady-state sensors, whereas, an extra blue outline marks the presence of unsteady sensors from a different run). The error bar marks the $12 \sigma$ deviation to highlight the less uncertainty in the measurements.}
\label{fig:sch_oil_pre}
\end{figure*}

Steady and unsteady sensors are mounted on the ramp-wall to monitor the flow separation and the associated flow oscillations. The location of those sensors between different runs are marked in Figures \ref{fig:schematic}, and \ref{fig:sch_oil_pre}. A 16-channel pressure scanner (Pressure Systems, Inc., Model-9016) is used to grab steady-state pressure data for a duration of 1 s at $f_s$=50 Hz. Unsteady pressure data is obtained using a fast-responding dynamic sensor (PCB\textsuperscript{\textregistered}, Model-113B24) for a period of 1 s at $f_s$=100 kHz. Hence, the spectral resolution and range is 1 Hz and 0-50 kHz.

Flat-frequency (20 to 16000 Hz) fast-responding array-microphones (PCB\textsuperscript{\textregistered}, Model-130A24) are used to measure the acoustic disturbances at a radius of $r=45L$ from the origin (see Figure \ref{fig:schematic}) at different angle ($\phi=45^\circ,90^\circ,$ and $135^\circ$) along the longitudinal plane ($xz$-plane) to the jet flow. The microphone samples data at $f_s=$0.4 MHz and the data is stored for a period of 2 s. Hence, the spectral resolution and range is 0.5 Hz and 0-0.2 MHz. A data acquisition card (NI-USB-6356) is used as an interface between the sensors and computers. A reference pressure of 20 $\mu$Pa is used for normalizing the premultiplied power spectra from the unsteady pressure and microphone measurements.

\subsection{Measurement uncertainty} \label{sec:uncertainty}

Each experiment is repeated three times to get a statistical consistency in the measured data. Uncertainties arising due to the repeatability, sensitivity, acquisition, data-conversion, and storage, and derived-data from image-processing are calculated as per the recommendations given in \cite{Coleman2009, Santo2004}. Uncertainty in identifying the spatial features from the oil flow and schlieren imaging is calculated to be 0.1 mm and 0.2 mm. \sk{The values are calculated based on the least pixel resolution during imaging.} The steady and unsteady pressure measurements have an uncertainty about the measured value of $\pm$3\% and $\pm$5\%, respectively. Microphone measurements have a total uncertainty of $\pm 5$ dB. The unsteady pressure and microphone data's spectral resolutions after zero-padding, windowing, and spectral smoothing are 0.5 Hz and 0.1 Hz. While using images from type-I imaging, the spatial features from the modal analysis comprise an uncertainty of about 0.2 mm. Whereas, the dominant and aliased spectra from the modal analysis have an uncertainty of about 1.25 Hz while using images from type-II imaging.

\section{Results and Discussions} \label{sec:res_disc}
\subsection{Prominent shrouded jet characteristics} \label{sec:jet_charac}

Preliminary observations from the high-speed schlieren at type-I imaging condition and oil flow visualization (see Figures \ref{fig:sch_oil_pre} I-II) show that the jet separates well before the ramp-wall termination. Time-averaged schlieren image (Figure \ref{fig:sch_oil_pre}-I) shows a jet deflection angle of $\delta \sim 12^\circ$ and $6^\circ$ about the streamwise direction ($x$-axis) for $\zeta=4$ and $5$, respectively. The shock-cells are closely packed and almost vanished earlier for lower $\zeta$ in comparison with the other. Instantaneous oil flow imaging (Figure \ref{fig:sch_oil_pre}-II) taken at the non-dimensional time step of $[t/T]=15$ and $25$ reveal the jet separation locations. For $\zeta=4$ and $5$, the separation is approximately seen at $[x/L] \sim 0.8$ and $0.7$. Jet with higher $p_0$ separates well ahead on the ramp-wall. \sk{When $p_0$ is large, the expansion angle at the top lip is even larger than the bottom lip (bottom lip has fixed expansion angle due to the ramp-wall). It tends to stretch the shock-cell further. The expansion fan from the bottom lip hits the top shear layer and turns back as compression waves. Later it becomes a strong shock wave. As the expansion angle in the top lip is larger for higher $p_0$, the turning compression waves become stronger. A stronger oblique shock hitting the ramp-wall produces a higher pressure ratio across the impinging point which is sufficient to separate the jet from the ramp-wall.}

The effects of the lateral expansion are evident in the oil flow visualization. At $\zeta=4$, the separation line curvature on the lateral side (marked as dotted-green line in Figure \ref{fig:sch_oil_pre} IIa-b) is larger in comparison to $\zeta=5$, where the separation line is almost fuller. For $\zeta=5$, only three shock-cells are sharply visible in contrast to the visibility of four shock-cells at $\zeta=4$ in a progressively smeared manner. \sk{The elongation of shock-cell to higher pressure ratio is due to the increment in the expansion fan-angle from the nozzle-lip. Such elongation will limit the viewing angle during imaging with limited optical-access which results in the observation of fewer shock-cells for higher $\zeta$.} 

In general, the first shock-cell spacing ($S_c/L$ at $c=1$) is found to be larger for all the cases of shrouded jets in comparison to the free jets as shown in Figure \ref{fig:sch_oil_pre}-III (see also Figure \ref{fig:xt_diagram} II-b) operating at the same $\zeta$ as per the model of \cite{Norum1982}. The model predicted the average shock-cell spacing as
\begin{equation}
    S_c = 1.1 \sqrt{M_j^2 - 1} \cdot D_e,
    \label{eq:cell_space}
\end{equation} where $D_e=\sqrt{4wh/\pi}$ is the equivalent diameter, $w$ and $h$ are the width and height of the jet nozzle. The shock-cells for $\zeta=5$ are not smeared and are considerably elongated (Figure \ref{fig:sch_oil_pre}-III), at least for the observed region in comparison to $\zeta=4$. \sk{In the free jet case, the expansion fan begins from both top and bottom lip in a symmetric manner. In the shrouded jet, the top lip let the expansion fan to expand freely until it balances out the pressure with the atmosphere ($p_a$). However, the bottom lip of the shrouded jet contains the ramp-wall with a fixed ramp angle of $15^\circ$. The expansion angle in the bottom lip is smaller than the top-lip. The smaller expansion angle takes longer distance to interact with the upper shear layer and thus elongating the first shock-cell.}

Figures \ref{fig:sch_oil_pre} I-II formulate the basis for steady and unsteady pressure sensors mounting, especially the spatial resolution based on the sensor form factor and the ramp-wall's structural rigidity to house a few of them. In Figure \ref{fig:sch_oil_pre}-IV, the non-dimensionalized time-averaged wall-static pressures are plotted along the streamwise direction. For both cases, a sudden pressure rise is observed at $[x/L] > 0.67$. However, in the upstream, owing to the higher $p_0$, case of larger $\zeta$ exhibit higher $[\overline{p}/p_a]$ on the ramp-wall. The exact position of the flow separation is difficult to identify due to poor resolution in sensor placement. Asides, the observations made in the oil flow measurements are consistent with the wall-static pressure measurements.

\subsection{Unsteady shock-cell oscillations and jet flapping}\label{sec:unsteady_jet_dyn}

\begin{figure*}
  \centerline{\includegraphics[width=0.8\textwidth]{ 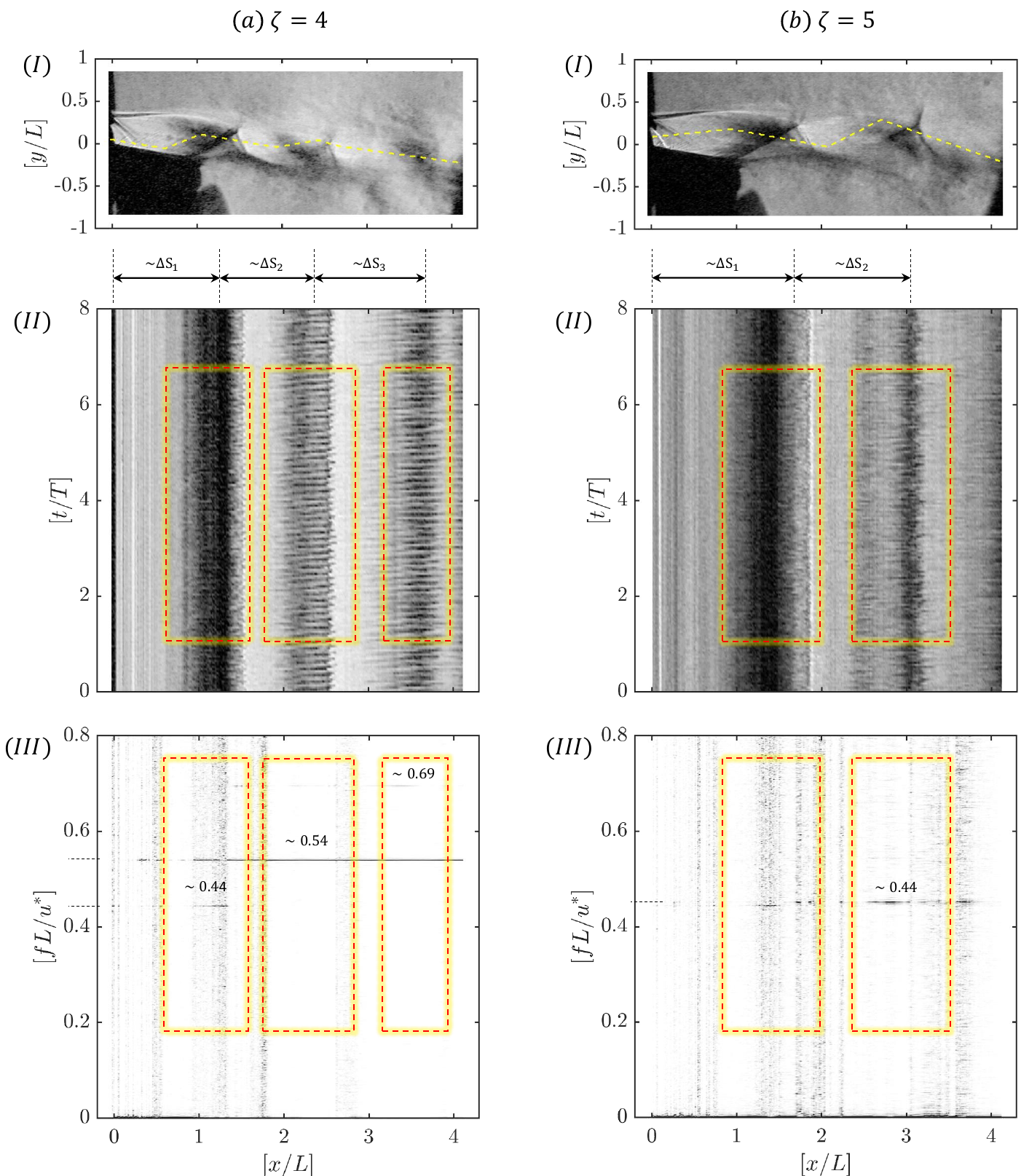}}
  \caption{Shock oscillation dynamics from the schlieren images of type-II imaging at $\zeta = 4, 5$ (a,b): I. Instantaneous schlieren imaging showing the line-profile (dotted-yellow line) along which the $x-t$ plot is constructed; II. $x-t$ plots showing the shock oscillation zone near the ramp-wall (dash-line red-box) and the periodic/non-periodic jet flapping downstream (dot-line red-box). \sk{Shock-cell spacing ($\Delta S_c$) is computed by taking the average spatial oscillation}; III. Non-dimensional normalized fast-Fourier transform (FFT) of the $x-t$ diagram (inverted-gray scale) revealing the dominant aliased spectral contents near and downstream the ramp-wall. \sk{Different ($\zeta=4$) and similar ($\zeta=5$) spectra} closer to the ramp-wall and downstream the nozzle are marked as dash-line red-box. Corresponding $x-t$ plot animation is available at \href{https://youtu.be/oaJCCeCWWyE}{`video3.mp4'} in the supplementary.}
\label{fig:xt_diagram}
\end{figure*}

The high-speed schlieren imaging done at type-II imaging condition reveals the unsteady features present in the jet. To evaluate the intrinsic dynamics closer to the ramp-wall and further downstream, the obtained instantaneous images are used to construct the $x-t$ diagram. \sk{A particular form of line profile is preferred to see the oscillation foot print clearly. A generic straight line along the core may not pass through the contrast shock regions exactly and will only print meagre oscillations that do not highlight the actual flow physics. Hence, in every instantaneous image, a consistent line profile shown as a dotted yellow line in Figure \ref{fig:xt_diagram}-I is drawn. The local light intensity in that profile is stacked along as time progress, and the $x-t$ plot is constructed as shown in Figure \ref{fig:xt_diagram}-II.} A fast-Fourier transform (FFT) is performed on the $x-t$ diagram to identify the spectral contents as shown in Figure \ref{fig:xt_diagram}-III. Corresponding $x-t$ plot animation is also available at \href{https://youtu.be/oaJCCeCWWyE}{`video3.mp4'} in the supplementary. \sk{It is worth to mention that the FFT of $x-t$ diagram will be completely different and unrealistic if the line profiles are taken at different locations. The line profiles for $\zeta=4$ and $5$ are non-identical for the aforementioned reason. The spectra from the microphones and unsteady pressure transducers help in arriving to a suitable line-profile for each of the cases.} 

\sk{The constructed $x-t$ diagram for $\zeta=4$ and $5$ (Figure \ref{fig:xt_diagram} IIa-b) reveals the presence of oscillating shocks.} The FFT analysis yields a spectra at an aliased non-dimensional frequency of $[fL/u^*]\sim 0.44$ and $[fL/u^*]\sim 0.69$, respectively. The oscillating shock's footprint is shown as dash-line red-box in Figure \ref{fig:xt_diagram}-II. Downstream the ramp-wall, the jet enters into the flapping mode. The shock-cell is smeared at portions where the jet-flap is vigorous. For $\zeta=4$, the jet flaps periodically at the end of the second shock-cell. The periodic pattern of the shock-cell footprint is apparent for $\zeta=4$ which is marked as dot-line red-box in Figure \ref{fig:xt_diagram}-II. The FFT contour plot reveals the presence of a narrow-band of frequency and its harmonics centered at $[fL/u^*]\sim 0.54$. Furthermore, the downstream jet oscillation for $\zeta=5$ looks non-periodic, as seen from the $x-t$ diagram. In those non-periodic regions, the FFT analysis reveals the presence of moderately broadened spectral contents centered at a non-dimensional frequency of $[fL/u^*]\sim 0.44$. The similar broadened frequency corresponding to the shock-induced separation on the wall at $[fL/u^*]\sim 0.44$ is also present in the downstream region, however at a strong intensity.

\subsection{Dominant spatiotemporal mode} \label{sec:spatial_mode}

\begin{figure*}
  \centerline{\includegraphics[width=0.8\textwidth]{ 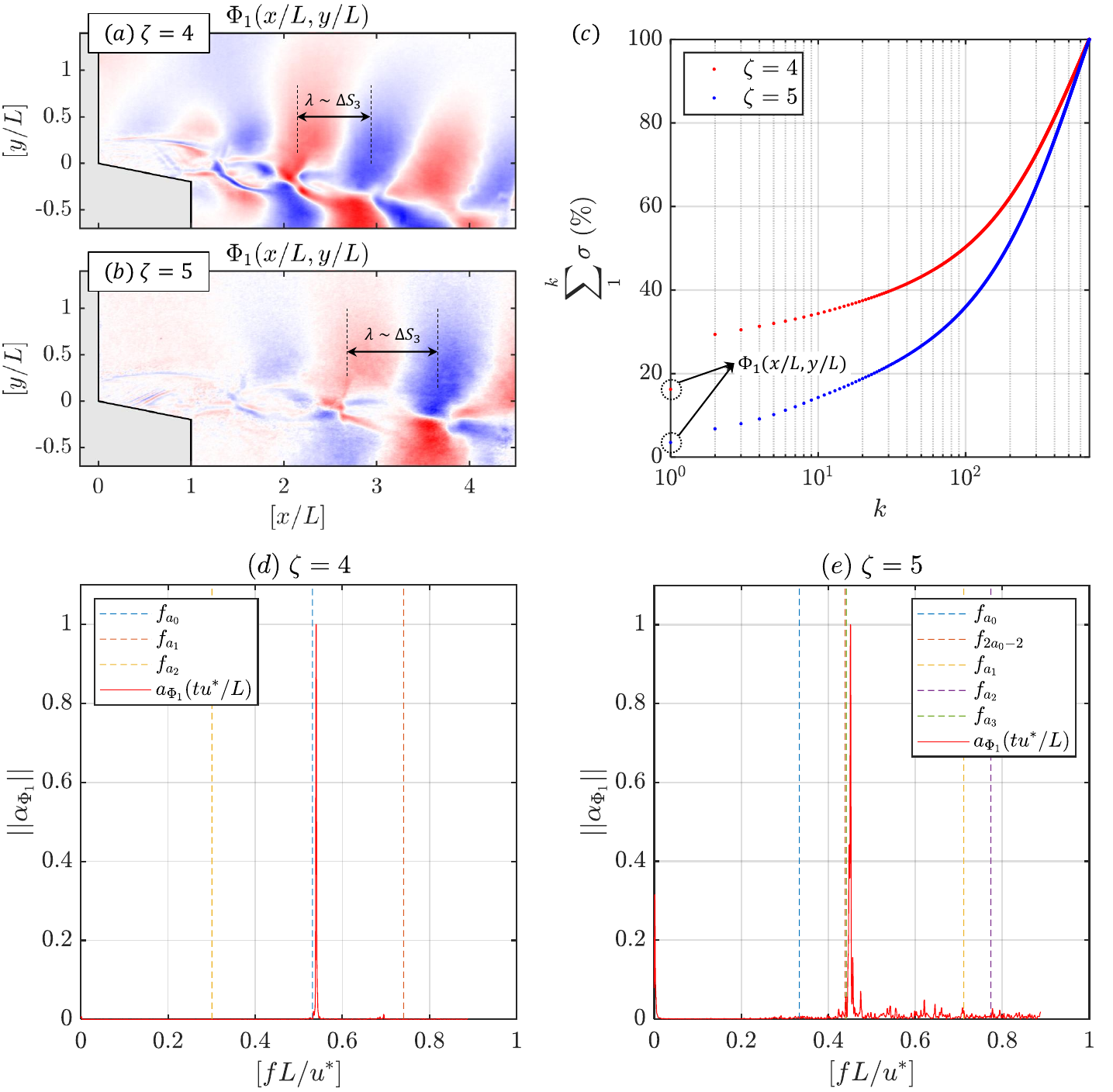}}
  \caption{(a-b). Normalized dominant energetic spatial mode from the POD analysis ($\Phi_1(x/L,y/L)$) for $\zeta =4$ and  (red and blue color contour correspond to +1 and -1 as equivalent to the positive and negative spatial correlation). \sk{The wavelength ($\lambda$) is quantified by measuring the distance between the color-map extrema intensities in a profile drawn along the jet shear layer;} (c). Cumulative energy distribution across the $k$th mode from the POD analysis for $\zeta =4$ and $5$; (d-e). Normalized non-dimensional spectra of the temporally varying POD coefficients ($a_{\Phi_1}(tu^*/L)$) corresponding to $\Phi_1(x/L,y/L)$ for $\zeta =4$ and $5$. The dotted-vertical lines in the subplots of d-e mark the different aliased frequencies ($f_a$) corresponding to the frequencies annotated in the microphone spectra shown in Figure \ref{fig:spectra} Ia-b.}
\label{fig:pod_diagrams}
\end{figure*}

The high-speed schlieren images are taken at two different frame rates and moderate light exposures (see $\S$\ref{sec:flow_vis}). However, as demonstrated by \cite{Rao2019}, extracting the flow dominant energetic mode from them is possible through one of the modal analysis techniques called Proper Orthogonal Decomposition-POD \citep{MEYER2007,Taira2017,Sahoo2021}. We took 700 instantaneous snapshots ($k=700$) for each case at type-I imaging condition and subjected them to the POD analysis as few of them are filtered due to imaging anomalies. The dominant energetic spatial mode [$\Phi_1(x/L,y/L)$] for $\zeta =4$ and $5$ is shown in Figure \ref{fig:pod_diagrams} a-b. The cumulative distribution of energy (square of the light intensity fluctuation in schlieren images) across the $k$th mode is plotted \sk{in Figure \ref{fig:pod_diagrams}c.}

From $\Phi_1(x/L,y/L)$ for $\zeta =4$ and $5$, the correlations pattern (alternate color patterns of red and blue) are observed to be anti-symmetric about the jet axis, corresponding to the sinuous mode of instability or simply jet-flapping \citep{Berland2007}. The strength of the correlation map is strong at the third shock-cell ($c=3$), at least based on the region of visualization interest. Furthermore, the spacing between them or the wavelength of the correlation map is also observed to be equivalent to the third shock-cell length ($\lambda \sim \Delta S_3$). The findings of $\lambda$ from $\Phi_1(x/L,y/L)$ is also matching closely with the relation proposed by \cite{Tam1988} as 
\begin{equation}
    \lambda = \frac{\Delta S_3}{n}\left(1+\frac{1}{M_c}\right) \\
\end{equation}
where $\lambda$ is the wavelength, $\Delta S_3$ is the shock-cell spacing, $n$ is the screech harmonic integer ($n=2$, reason is explained in $\S$\ref{sec:shock_acous}), and $M_c$ is the convective Mach number.

POD based modal analysis shows that the first mode for $\zeta =4$ and $5$ contain about $\sim 16$\% and $\sim 4$\% of the total energy in the first mode itself. The strongly correlated spatial field and larger energy content in the first mode ($k=1$) for $\zeta=4$ indicate the dominant form of jet-screech in comparison with the other one. For lower frame rate cases like the present one, investigating other spatiotemporal modes than the dominant one is reportedly difficult without supplementary measurements \cite{Rao2019,Rao2020}. In Figure \ref{fig:pod_diagrams}d-e images from the type-II imaging conditions are used and the FFT spectra of time varying POD coefficients ($a_{\Phi_1}(tu^*/L)$) corresponding to the dominant mode is presented. The spectra is aliased except for the dominant one and the values need to be interpretted with the fully resolved relevant measurements like the microphone or unsteady pressure. However, from the cursory investigation, the dominant spectra for $\zeta=4$ looks narrower and concentrated at $[fL/u^*] \sim 0.55$. Likewise, the dominant spectra for $\zeta=5$ is observed at $[fL/u^*] \sim 0.44$ but looks slightly broadened.

\subsection{Separation unsteadiness and jet aeroacoustics} \label{sec:shock_acous}
Preliminary acoustic measurements in Figure \ref{fig:run_time} are analyzed first to access the generic difference in the tonal signature of the shrouded jets between different $\zeta$. \sk{However, as a cautionary note, it has to mentioned that the microphone data has not been taken at high directional resolution ($d\phi$) as only three longitudinal locations ($\phi$) are probed due to limitations in deploying many sensors, appropriate mounts, and data acquisition cards. Hence, the drawn conclusions are only limited to the respective $\phi$ but not everywhere or around it}. At $\zeta=4$ \sk{(Figure \ref{fig:run_time}a)}, a fundamental and its first harmonic are seen at $[fL/u^*]\sim 0.55$ and $\sim 1.1$. As $p_0$ increases, the jet at $\zeta=5$ \sk{(Figure \ref{fig:run_time}b)} undergoes mode switching \citep{Chen_2018}. The fundamental tone switches between $[fL/u^*]\sim 0.47$ and $\sim 0.37$ along with its harmonic for few seconds and then remains steady. The shrouded jet's dominant tone obtained during the steady-state is compared with the free jet using the relation of \cite{Tam1988} as shown in Eq. \ref{eq:st_var}. For the larger aspect ratio free jet flows with a specific heat capacity of $\gamma=1.4$, the non-dimensional frequency for different $M_j$ can be calculated as
\begin{equation}
\begin{split}
\displaystyle \left[\frac{fh_j}{u_j}\right] & \approx \left[{0.7 M_j \left( \displaystyle \frac{1.2}{1+0.2M_j^2}\right)^3}\right] \times\\ &\left\{2(M_j^2-1) \displaystyle \left[1 + \frac{0.7M_j}{\left(1+0.2M_j^2\right)^{1/2}}\right]\right\}^{-1}.
\label{eq:st_var}
\end{split}
\end{equation}
While comparing the free jet data with that of the shrouded jet, the non-dimensional frequency ($fh_j/u_j$) for $\zeta=4$ and $5$ are found to be higher than the predicted values from Eq. \ref{eq:st_var} by 13-19 \%.

\begin{figure*}
  \centerline{\includegraphics[width=0.75\textwidth]{ 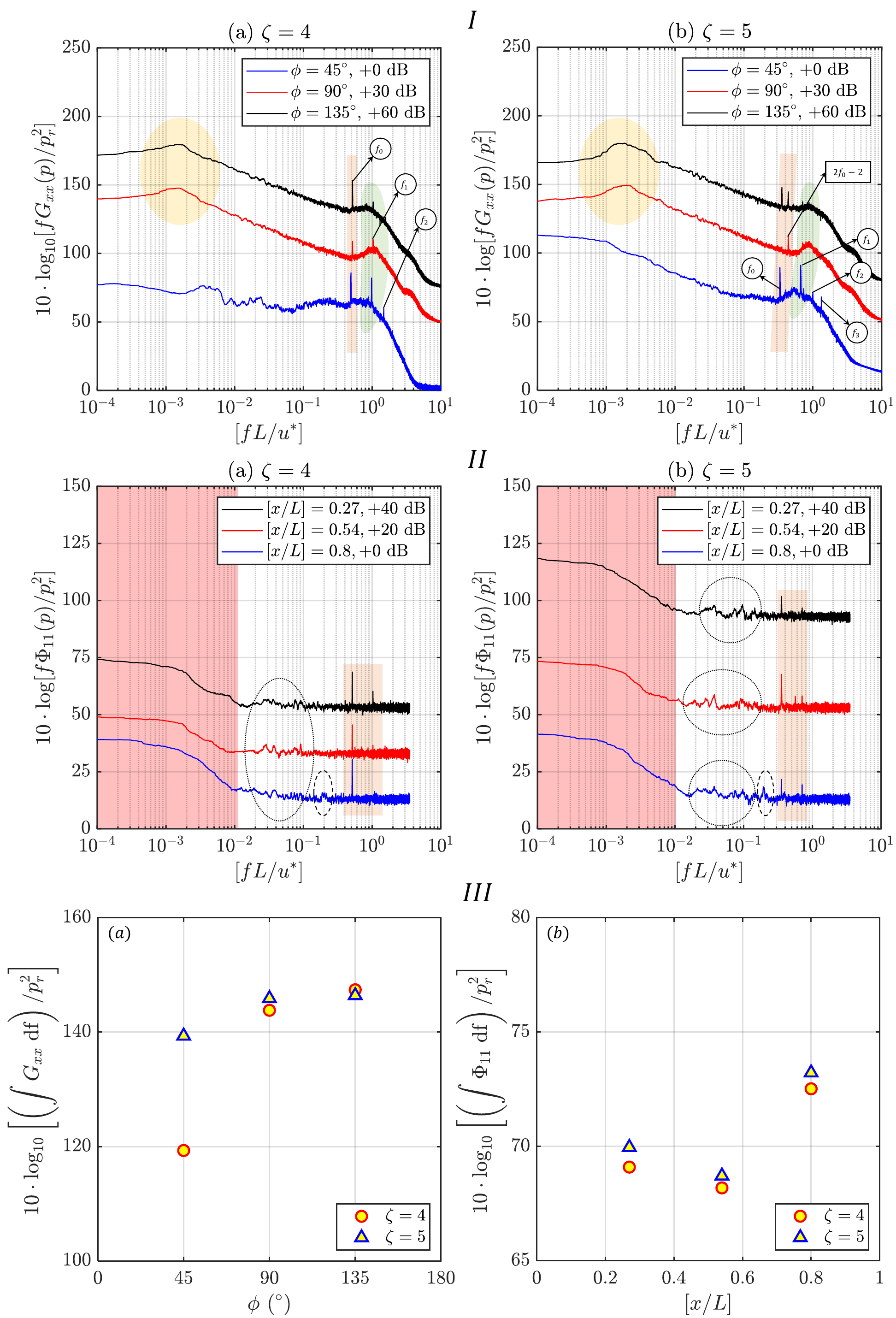}}
  \caption{I. Microphone spectra obtained for $\zeta =4,5$ (a, b) at different $\phi$ ($^\circ$) as shown in Figure \ref{fig:schematic}. Some of the dominant frequencies are numbered for discussion purpose. The shaded regions in the plots denote the following: yellow-turbulent mixing noise, orange-jet screeching, green-broadband shock-associated noise; II. Unsteady pressure spectra from the three sensors mounted on the planar-plug as shown in Figures \ref{fig:schematic} and \ref{fig:sch_oil_pre}. Dotted line: low-frequency shock-related unsteadiness, Dash line: moderate frequency unsteadiness related to shock-induced separation, Orange shade: pressure fluctuations associated with the acoustic forcing from jet-screech, Red shade: marks the cut-off frequency of 200 Hz. \sk{III. Overall energy contents from the (a) microphone and (b) pressure spectra at different $\phi$ ($^\circ$) and $[x/L]$ (red-outlined circles for $\zeta=4$ and blue-outline triangles for $\zeta=5$).}}
\label{fig:spectra}
\end{figure*}

Detailed microphone measurements are taken at different $\phi$ in an semi anechoic chamber to reveal the presence of dominant tonal noise for both cases. In all the directions as shown in Figure \ref{fig:spectra}-Ia, $\zeta=4$ case exhibit a tonal noise having the fundamental at $[f_0 L/u^*] \sim 0.5$ and their two overtones at ($[f_1 L/u^*] \sim 1$ and $[f_2 L/u^*] \sim 1.5$), respectively. The fundamental being the dominant one producing a sound pressure level of about 90 dB. The uniform radiation of sound across all the directions denotes the existence of strong screeching jet behavior. Asides jet-screech (tonal frequencies marked as orange shade in Figure \ref{fig:spectra}-Ia), the other noise components like the turbulent mixing noise (yellow shade) is only prominent in the far-field for $\phi=90^\circ$ and $\phi=135^\circ$ to the jet axis with a peak noise of $\sim 120$ dB at $[f_1 L/u^*] \sim 2 \times 10^{-3}$. For $\phi=45^\circ$, turbulent-mixing noise is suppressed to $\sim 77$ dB and observed at $[f_1 L/u^*] \sim 4 \times 10^{-3}$. The suppression could be attributed to the screech dominance. The broadband shock associated noise (BSAN, shown as green shade in Figure \ref{fig:spectra}-Ia) is dominant only at $\phi=90^\circ$ with a maximum sound pressure level of $\sim 80$ dB ($[f_1 L/u^*] \sim 1$) and felt considerably at other directions also.

On the other-hand, for $\zeta=5$, the fundamental ($[f_0 L/u^*] \sim 0.3$) and the first overtone ($[f_1 L/u^*] \sim 0.6$) are almost comparable in sound pressure level of about 85 dB only for $\phi=45^\circ$. The second and third overtones are also present, however at a lower intensity. It has to be noted that these tonal noise are not consistent at other $\phi$. For $\phi=90^\circ$, BSAN at $[fL/u^*]\sim 0.8$ and a tonal noise at $[2f_0-2][L/u^*]\sim 0.4$ are dominant with a sound pressure level of $\sim 80$ dB. For $\phi=135^\circ$, the fundamental, intermediate, first overtone and the BSAN frequencies are present with the fundamental being the dominant one (at $\sim 87$ dB). The turbulent mixing noise is also observed for $\phi=90^\circ$ and $135^\circ$ at $[f L/u^*] \sim 2 \times 10^{-3}$ with a sound pressure level of $\sim120$ dB.

\sk{The overall sound pressure level (OASPL, dB) from the microphone measurements and the total energy contents from the pressure fluctuations through unsteady pressure transducers at different operating conditions are shown in Figure \ref{fig:spectra}-III. At $\zeta=4$ (Figure \ref{fig:spectra}-IIIa), as $\phi$ increases, OASPL increases drastically from $\phi=45^\circ$ to $90^\circ$ and gradually between $90^\circ$ to $135^\circ$. For $\zeta=5$, OASPL on the jet stream side like $\phi=45^\circ$ is observed to be almost $17\%$ larger than the other. Away from the jet stream side, OASPL variations are not significant. The streamwise sound radiation (Mach wave radiation \cite{Tam2009}) from the turbulent structures convecting at comparatively higher velocities is attributed to the higher OASPL observation for $\zeta=5$. Moreover, at $\zeta=4$, the jet flaps severely as seen from the dominant spatial mode (Figure \ref{fig:pod_diagrams}a-b) and high-speed schlieren images (Figure \ref{fig:xt_diagram}). The overall energy contents from the unsteady pressure spectra (Figure \ref{fig:spectra}-IIIb) reveal the presence of comparatively higher magnitude for $\zeta=5$ than $\zeta=4$ everywhere ($x/L=0.27,0.54,0.8$). The higher pressure ratio difference or the shock strength for $\zeta=5$ and higher receptivity from the acoustic forcing by the screeching jet lead to the aforementioned observation.}

In summary, the turbulent mixing noise is perceived to be of the same intensity except spectral broadening at $\zeta=5$ for $\phi=90^\circ$ and $135^\circ$. The delayed jet separation ($x/L \sim 0.8$) and larger jet deflection ($\delta \sim 12^\circ$), keeps the jet closer to the ramp-wall, thereby reduce the intensity of sound emission at $\phi=45^\circ$ for $\zeta=4$. The lifting jet ($\delta \sim 6^\circ$) radiates turbulent mixing noise more, and hence a considerable amount of sound pressure level observed at $\phi=135^\circ$ for $\zeta=5$. An intermediate tonal noise around $[2f_0-2]$ for $\zeta=5$ at $\phi=90^\circ$ and $135^\circ$ is seen. The jet's lateral expansion turns most of the flow inwards (see Figure \ref{fig:sch_oil_pre} II-b), and the lifted jet entrains more fluid from the ambient. The resulting mixing and shock interactions might produce a lateral jet oscillation, and the intermediate tonal noise could be attributed to it.

The modal analysis is performed on a under-sampled schlieren images. Hence, the POD temporal coefficients are aliased in the spectra. The dominant spectral components corresponding to the jet-screech are identified from the microphone measurements. The aliased frequencies ($f_a$) in the dominant mode POD spectra are identified by using the formula 
\begin{equation}
    f_a = \left \| f - f_s \left \lfloor \frac{f}{f_s} \right \rceil \right \|
    \label{eq:alias}
\end{equation} 
where $f$ is the actual frequency, and $f_s$ is the frequency of the under-sampled data. By using Eq. \ref{eq:alias}, the possible dominant energetic mode and the corresponding non-dimensional frequency can be identified. Based on the $\lambda$, and $u_c$, the frequencies are computed for both the $\zeta=4$ and $5$. They are identified to be almost equal to the second harmonic ($n=2$) of the jet-screech. In Figure \ref{fig:pod_diagrams} d-e, the aliased microphone frequencies are projected and they are found to be fairly matching with the POD spectra corresponding to the dominant energetic mode.

In addition, such behaviour is observed only at restricted direction of $\phi=45^\circ$ and $135^\circ$. All flapping-jets exhibiting sinuous mode, in general show such aeroacoustics. At $\phi=45^\circ$, a discrete tone between $f_0$ and $f_1$ is observed ($[2f_0-2] [L/u^*] \sim 0.4$). The absence of uniform tonal dominance in all the direction indicates the absence of the general of spherical/cylindrical acoustic disturbances from the screeching jet. From the run-time spectrograms shown in Figure \ref{fig:run_time}, mode switching is seen for $\zeta=5$, indicating that the jet has ventured into the new zone of screeching mode at $\zeta$ increases.

Simultaneously, to monitor the separation unsteadiness, three unsteady pressure sensors are mounted on the ramp-wall. The spectral signature arising from the unsteady pressure sensor (see Figure \ref{fig:spectra} IIa-b) and the microphone measurements (see Figure \ref{fig:spectra} Ia-b) are fairly matching. Especially on the sensors closer to the point of separation ($[x/L] \sim 0.8$ for $\zeta=4$ and $[x/L] \sim 0.54$ for $\zeta=5$), the larger power density depict the stronger correlation between the emitted acoustics and the flow oscillation. The generated acoustic waves for both the cases affect the upper shear layer directly and the subsonic part of the boundary layer on the ramp-wall through acoustic forcing. Such an event forms closed feedback and result in the self-sustained oscillation of the jet.

In Figure \ref{fig:spectra} IIa-b, the shock-related unsteadiness on the ramp-wall are confined to the lower end of the spectrum ($10^{-2} \leq [fL/u^*] \leq 10^{-1}$). The shock-induced jet separation produces a broadened spectra in the middle ($fL/u^* \sim 2 \times 10^{-1}$). The spectra are in general seen only at $[x/L]=0.8$ with different amplitudes for both $\zeta =4$ (less) and $5$ (high), as that particular sensor is closer to the point of jet-separation. Jet-screech induced acoustic forcing is seen throughout the positioned three sensors on the ramp-wall. Although the ramp-wall shrouds the bottom portion of the jet, the top-shear layer's acoustic forcing distorts the shocks inside the jet, thereby altering the shock-foot print on the ramp-wall. 

This effect is well captured in the dominant energetic spatial mode in Figure \ref{fig:pod_diagrams}a-b. The correlation map is persistent on the top and bottom portion of the jet. However, the contour map intensities are stronger for $\zeta=4$ ,which means that the acoustic forcing is severe. Furthermore, the premature separation in $\zeta=5$ and the jet-flap happening at the end of the third or fourth shock-cell considerably reduces the acoustic forcing on this particular jet. The drop in jet-screech forcing on the jet-separation region is at $[x/L]=0.8$ unsteady pressure spectra (orange shaded region in Figure \ref{fig:spectra} IIb). 

\section{Conclusions} \label{sec:conclusions}

A partially shrouded compressible jet exhausting from a planar-plug nozzle operating at two nozzle pressure ratio of $\zeta=4$ and $5$ is considered to study the unsteady flow dynamics and the resulting aeroacoustics. Following are the major findings:
\begin{itemize}
    \item {The schlieren and oil flow visualization shows that the jet deflects away from the ramp-wall more and separate ahead at high $\zeta$.}
    \item {The unsteady pressure data along the ramp wall show the existence of distinct tonal disturbances responsible for oscillations closer to flow separation.}
    \item {The microphone measurements also found to be following the trends of unsteady spectra in terms of distinct tonal disturbances. At $\zeta=4$, the jet screeches at a larger intensity in comparison with the other.}
    \item {A strong coupling between the unsteady flow mode and the dominant acoustic mode is seen for both cases.}
    \item {\sk{The acoustic disturbances from the flapping jet force the upper shear layer and the subsonic boundary layer on the ramp-wall about the separation point. The periodic forcing further introduces instability in the jet shear layer which in-turn cause the jet flapping. Thus, the established closed feedback loop results in the self-sustained jet oscillation and sound production.}}
\end{itemize}

\section*{Supplemental material}
See supplemental material for a video showing the schlieren images from type-I imaging condition under the name \href{https://youtu.be/ZN-0Ah19Pwc}{`video1.mp4'}. See supplemental material for a video showing the oil flow visualization and the corresponding microphone spectra under the name \href{https://youtu.be/1t0hw1pHnLw}{`video2.mp4'}. See supplemental material for a video showing the $x-t$ plot animation from type-II imaging condition under the name \href{https://youtu.be/oaJCCeCWWyE}{`video3.mp4'}.

\section*{Authors' Contributions}
S.R.N., S.K.K., T.V.K., A.D., and I.M.S. have contributed equally to this work.

\section*{Acknowledgement}

The financial support from the Department of Aerospace Engineering, Indian Institute of Technology, Kanpur-India, and DST-FIST, India, for carrying out the research work is gratefully acknowledged. The data that support the findings of this study are available from the corresponding author upon reasonable request. All authors have contributed equally to this work. The authors report no conflicts of interest.  

\section*{Data availability statement}
The data that support the findings of this study are available from the corresponding author upon reasonable request.


\bibliographystyle{apalike}
\bibliography{references}

\begin{thebibliography}{}

\bibitem[Baskaran and Srinivasan, 2019]{Baskaran2019}
Baskaran, K. and Srinivasan, K. (2019).
\newblock Effects of upstream pipe length on pipe-cavity jet noise.
\newblock {\em Physics of Fluids}, 31(10):106103.

\bibitem[Behrouzi et~al., 2018]{Behrouzi2018}
Behrouzi, P., McGuirk, J.~J., and Avenell, C. (2018).
\newblock Effect of scarfing on rectangular nozzle supersonic jet plume flow
  characteristics.
\newblock {\em {AIAA} Journal}, 56(1):301--315.

\bibitem[Berland et~al., 2007]{Berland2007}
Berland, J., Bogey, C., and Bailly, C. (2007).
\newblock Numerical study of screech generation in a planar supersonic jet.
\newblock {\em Physics of Fluids}, 19(7):075105.

\bibitem[Berry et~al., 2017a]{Berry2017c}
Berry, M.~G., Ali, M.~Y., Magstadt, A.~S., and Glauser, M.~N. (2017a).
\newblock {DMD} and {POD} of time-resolved schlieren on a multi-stream single
  expansion ramp nozzle.
\newblock {\em International Journal of Heat and Fluid Flow}, 66:60--69.

\bibitem[Berry et~al., 2017b]{Berry2017b}
Berry, M.~G., Magstadt, A.~S., and Glauser, M.~N. (2017b).
\newblock Application of {POD} on time-resolved schlieren in supersonic
  multi-stream rectangular jets.
\newblock {\em Physics of Fluids}, 29(2):020706.

\bibitem[Berry et~al., 2017c]{Berry2017a}
Berry, M.~G., Stack, C.~M., Magstadt, A.~S., Ali, M.~Y., Gaitonde, D.~V., and
  Glauser, M.~N. (2017c).
\newblock Low-dimensional and data fusion techniques applied to a supersonic
  multistream single expansion ramp nozzle.
\newblock {\em Physical Review Fluids}, 2(10).

\bibitem[Chaudhary et~al., 2020]{Chaudhary2020}
Chaudhary, M., Krishna, T.~V., Nanda, S.~R., Karthick, S.~K., Khan, A., De, A.,
  and Sugarno, I.~M. (2020).
\newblock On the fluidic behavior of an over-expanded planar plug nozzle under
  lateral confinement.
\newblock {\em Physics of Fluids}, 32(8):086106.

\bibitem[Chen et~al., 2018]{Chen_2018}
Chen, Z., Wu, J.-H., Ren, A.-D., and Chen, X. (2018).
\newblock Mode-switching and nonlinear effects in supersonic jet noise.
\newblock {\em {AIP} Advances}, 8(1):015126.

\bibitem[Chutkey et~al., 2012]{Chutkey_2012}
Chutkey, K., Vasudevan, B., and Balakrishnan, N. (2012).
\newblock Flowfield analysis of linear plug nozzle.
\newblock {\em Journal of Spacecraft and Rockets}, 49(6):1109--1119.

\bibitem[Chutkey et~al., 2017]{Chutkey_2017}
Chutkey, K., Viji, M., and Verma, S.~B. (2017).
\newblock Effect of clustering on linear plug nozzle flow field for
  overexpanded internal jet.
\newblock {\em Shock Waves}, 27(4):623--633.

\bibitem[Chutkey et~al., 2018]{Chutkey_2018}
Chutkey, K., Viji, M., and Verma, S.~B. (2018).
\newblock Interaction of external flow with linear cluster plug nozzle jet.
\newblock {\em Shock Waves}, 28(6):1207--1221.

\bibitem[Clemens and Narayanaswamy, 2014]{Clemens2014}
Clemens, N.~T. and Narayanaswamy, V. (2014).
\newblock Low-frequency unsteadiness of shock wave/turbulent boundary layer
  interactions.
\newblock {\em Annual Review of Fluid Mechanics}, 46(1):469--492.

\bibitem[Coleman and Steele, 2009]{Coleman2009}
Coleman, H.~W. and Steele, W.~G. (2009).
\newblock {\em Experimentation, Validation, and Uncertainty Analysis for
  Engineers}.
\newblock John Wiley {\&} Sons, Inc.

\bibitem[Das and Dosanjh, 1991]{Das1991}
Das, I. and Dosanjh, D. (1991).
\newblock Short conical solid/perforated plug-nozzle as supersonic jet noise
  supressor.
\newblock {\em Journal of Sound and Vibration}, 146(3):391--406.

\bibitem[Das et~al., 1997]{Das1997}
Das, I., Khavaran, A., and Krejsa, E. (1997).
\newblock A computational study of contoured plug-nozzle jet noise.
\newblock {\em Journal of Sound and Vibration}, 206(2):169--194.

\bibitem[Dhamanekar and Srinivasan, 2013]{Dhamanekar2013}
Dhamanekar, A. and Srinivasan, K. (2013).
\newblock Hysteresis effects in the impinging jet noise.
\newblock {ASA}.

\bibitem[Dosanjh and Das, 1988]{Dosanjh1988}
Dosanjh, D.~S. and Das, I.~S. (1988).
\newblock Aeroacoustics of supersonic jet flows from a contoured plug-nozzle.
\newblock {\em {AIAA} Journal}, 26(8):924--931.

\bibitem[Edgington-Mitchell, 2019]{EdgingtonMitchell2019}
Edgington-Mitchell, D. (2019).
\newblock Aeroacoustic resonance and self-excitation in screeching and
  impinging supersonic jets {\textendash} a review.
\newblock {\em International Journal of Aeroacoustics}, 18(2-3):118--188.

\bibitem[Estruch-Samper and Chandola, 2018]{EstruchSamper2018}
Estruch-Samper, D. and Chandola, G. (2018).
\newblock Separated shear layer effect on shock-wave/turbulent-boundary-layer
  interaction~unsteadiness.
\newblock {\em Journal of Fluid Mechanics}, 848:154--192.

\bibitem[Hiley et~al., 1976]{Hiley1976}
Hiley, P.~E., Wallace, H.~W., and Booz, D.~E. (1976).
\newblock Nonaxisymmetric nozzles installed in advanced fighter aircraft.
\newblock {\em Journal of Aircraft}, 13(12):1000--1006.

\bibitem[Karthick et~al., 2016]{Karthick2016}
Karthick, S.~K., Rao, S. M.~V., Jagadeesh, G., and Reddy, K. P.~J. (2016).
\newblock Parametric experimental studies on mixing characteristics within a
  low area ratio rectangular supersonic gaseous ejector.
\newblock {\em Physics of Fluids}, 28(7):076101.

\bibitem[Khan et~al., 2019]{Khan_2019}
Khan, A., Panthi, R., Kumar, R., and Ibrahim, S.~M. (2019).
\newblock Experimental investigation of the effect of extended cowl on the flow
  field of planar plug nozzles.
\newblock {\em Aerospace Science and Technology}, 88:208--221.

\bibitem[Li et~al., 2019]{Li2019}
Li, Y., He, C., Li, J., Miao, L., Gao, R., and Liang, J. (2019).
\newblock Experimental investigation of flow separation in a planar
  convergent-divergent nozzle.
\newblock {\em Journal of Physics: Conference Series}, 1300:012088.

\bibitem[Malla and Gutmark, 2021]{Malla2021}
Malla, B. and Gutmark, E.~J. (2021).
\newblock Single expansion ramp nozzles: Impact of ramp length on flow and
  acoustics.
\newblock {\em {AIAA} Journal}, pages 1--13.

\bibitem[Meyer et~al., 2007]{MEYER2007}
Meyer, K.~E., Pedersen, J.~M., and \"{O}zcan, O. (2007).
\newblock A turbulent jet in crossflow analysed with proper orthogonal
  decomposition.
\newblock {\em Journal of Fluid Mechanics}, 583:199--227.

\bibitem[Norum and Seiner, 1982]{Norum1982}
Norum, T.~D. and Seiner, J.~M. (1982).
\newblock Broadband shock noise from supersonic jets.
\newblock {\em {AIAA} Journal}, 20(1):68--73.

\bibitem[Panda et~al., 1997]{Panda1997}
Panda, J., Raman, G., Zaman, K., Panda, J., Raman, G., and Zaman, K. (1997).
\newblock Underexpanded screeching jets from circular, rectangular and elliptic
  nozzles.
\newblock In {\em 3rd {AIAA}/{CEAS} Aeroacoustics Conference}. American
  Institute of Aeronautics and Astronautics.

\bibitem[Rao and Karthick, 2019]{Rao2019}
Rao, S.~M. and Karthick, S. (2019).
\newblock Studies on the effect of imaging parameters on dynamic mode
  decomposition of time-resolved schlieren flow images.
\newblock {\em Aerospace Science and Technology}, 88:136--146.

\bibitem[Rao and Jagadeesh, 2014]{Rao2014}
Rao, S. M.~V. and Jagadeesh, G. (2014).
\newblock Observations on the non-mixed length and unsteady shock motion in a
  two dimensional supersonic ejector.
\newblock {\em Physics of Fluids}, 26(3):036103.

\bibitem[Rao et~al., 2020]{Rao2020}
Rao, S. M.~V., Karthick, S.~K., and Anand, A. (2020).
\newblock Elliptic supersonic jet morphology manipulation using sharp-tipped
  lobes.
\newblock {\em Physics of Fluids}, 32(8):086107.

\bibitem[Romine, 1998]{Romine1998}
Romine, G.~L. (1998).
\newblock Nozzle flow separation.
\newblock {\em {AIAA} Journal}, 36(9):1618--1625.

\bibitem[Sahoo et~al., 2020]{Sahoo2020}
Sahoo, D., Karthick, S.~K., Das, S., and Cohen, J. (2020).
\newblock Parametric experimental studies on supersonic flow unsteadiness over
  a hemispherical spiked body.
\newblock {\em {AIAA} Journal}, 58(8):3446--3463.

\bibitem[Sahoo et~al., 2021]{Sahoo2021}
Sahoo, D., Karthick, S.~K., Das, S., and Cohen, J. (2021).
\newblock Shock-related unsteadiness in spiked-body flow at supersonic speed.
\newblock {\em Experiments in Fluids}, 58(8):3446--3463.

\bibitem[Santo et~al., 2004]{Santo2004}
Santo, M.~D., Liguori, C., Paolillo, A., and Pietrosanto, A. (2004).
\newblock Standard uncertainty evaluation in image-based measurements.
\newblock {\em Measurement}, 36(3-4):347--358.

\bibitem[Sethuraman et~al., 2021]{PethaSethuraman2021}
Sethuraman, V. R.~P., Kim, T.~H., and Kim, H.~D. (2021).
\newblock Effects of back pressure perturbation on shock train oscillations in
  a rectangular duct.
\newblock {\em Acta Astronautica}, 179:525--535.

\bibitem[Settles, 2001]{Settles2001}
Settles, G.~S. (2001).
\newblock {\em Schlieren and Shadowgraph Techniques}.
\newblock Springer Berlin Heidelberg.

\bibitem[Stack and Gaitonde, 2018]{Stack2018}
Stack, C.~M. and Gaitonde, D.~V. (2018).
\newblock Shear layer dynamics in a supersonic rectangular multistream nozzle
  with an aft-deck.
\newblock {\em {AIAA} Journal}, 56(11):4348--4360.

\bibitem[Sushil et~al., 2020]{Sushil2020}
Sushil, S.~K., Garg, M., and Narayanan, S. (2020).
\newblock Estimation of the lower cut-off frequency of an anechoic chamber: An
  empirical approach.
\newblock {\em International Journal of Aeroacoustics}, 19(1-2):57--72.

\bibitem[Sutton, 2006]{Sutton2006}
Sutton, G.~P. (2006).
\newblock {\em History of Liquid Propellant Rocket Engines}.
\newblock American Institute of Aeronautics and Astronautics.

\bibitem[Taira et~al., 2017]{Taira2017}
Taira, K., Brunton, S.~L., Dawson, S. T.~M., Rowley, C.~W., Colonius, T.,
  McKeon, B.~J., Schmidt, O.~T., Gordeyev, S., Theofilis, V., and Ukeiley,
  L.~S. (2017).
\newblock Modal analysis of fluid flows: An overview.
\newblock {\em {AIAA} Journal}, 55(12):4013--4041.

\bibitem[Tam, 1988]{Tam1988}
Tam, C. (1988).
\newblock The shock-cell structures and screech tone frequencies of rectangular
  and non-axisymmetric supersonic jets.
\newblock {\em Journal of Sound and Vibration}, 121(1):135--147.

\bibitem[Tam, 2009]{Tam2009}
Tam, C. K.~W. (2009).
\newblock Mach wave radiation from high-speed jets.
\newblock {\em {AIAA} Journal}, 47(10):2440--2448.

\bibitem[Terzis et~al., 2011]{Terzis2011}
Terzis, A., Zachos, P., Charnley, B., Pachidis, V., and Kalfas, A.~I. (2011).
\newblock On the applicability of oil and dye flow visualization technique
  during the design phase and operation of experimental rigs.
\newblock {\em Journal of Flow Visualization and Image Processing},
  18(3):199--214.

\bibitem[Veen et~al., 1974]{Veen1974}
Veen, R.~V., Gentry, R., and Hoffman, J.~D. (1974).
\newblock Design of shrouded-plug nozzles for maximum thrust.
\newblock {\em {AIAA} Journal}, 12(9):1193--1197.

\bibitem[Verma and Viji, 2011]{Verma_2011}
Verma, S. and Viji, M. (2011).
\newblock Linear-plug flowfield and base pressure development in freestream
  flow.
\newblock {\em Journal of Propulsion and Power}, 27(6):1247--1258.

\bibitem[Viswanathan and Czech, 2011]{Viswanathan2011}
Viswanathan, K. and Czech, M.~J. (2011).
\newblock Adaptation of the beveled nozzle for high-speed jet noise reduction.
\newblock {\em {AIAA} Journal}, 49(5):932--944.

\bibitem[Wei et~al., 2019]{Wei2019}
Wei, X., Mariani, R., Chua, L., Lim, H., Lu, Z., Cui, Y., and New, T. (2019).
\newblock Mitigation of under-expanded supersonic jet noise through stepped
  nozzles.
\newblock {\em Journal of Sound and Vibration}, 459:114875.

\bibitem[Wlezien and Kibens, 1988]{Wlezien1988}
Wlezien, R.~W. and Kibens, V. (1988).
\newblock Influence of nozzle asymmetry on supersonic jets.
\newblock {\em {AIAA} Journal}, 26(1):27--33.

\bibitem[Zaman, 1996]{Zaman1996}
Zaman, K. B. M.~Q. (1996).
\newblock Axis switching and spreading of an asymmetric jet: the role of
  coherent structure dynamics.
\newblock {\em Journal of Fluid Mechanics}, 316:1--27.

\bibitem[Zeoli and Gu, 2008]{Zeoli2008}
Zeoli, N. and Gu, S. (2008).
\newblock Computational validation of an isentropic plug nozzle design for gas
  atomisation.
\newblock {\em Computational Materials Science}, 42(2):245--258.

\bibitem[Zhu et~al., 2020]{Zhu2020}
Zhu, Y., Wang, K., Wang, Z., Zhao, M., Jiao, Z., Wang, Y., and Fan, W. (2020).
\newblock Study on the performance of a rotating detonation chamber with
  different aerospike nozzles.
\newblock {\em Aerospace Science and Technology}, 107:106338.

\end{thebibliography}

\onecolumngrid

\PRLsep	

\end{document}